\documentclass[fleqn,usenatbib]{mnras}

\usepackage[T1]{fontenc}
\usepackage{hyperref}
\usepackage[flushleft]{threeparttable}

\DeclareRobustCommand{\VAN}[3]{#2}
\let\VANthebibliography\thebibliography
\def\thebibliography{\DeclareRobustCommand{\VAN}[3]{##3}\VANthebibliography}

\usepackage{graphicx}	
\usepackage{amsmath}	
\usepackage{amssymb}	
\usepackage{siunitx}
\usepackage{float}
\usepackage{multirow}
\usepackage{placeins}
\usepackage{soul}
\usepackage{longtable}
\usepackage{subfigure}
\usepackage[table,x11names]{xcolor}
\usepackage{lscape}
\usepackage{comment}

\usepackage{newtxtext,newtxmath}

\title[The Relative Supernovae Contribution to the Chemical Enrichment History of Abell 1837]{The Relative Supernovae Contribution to the Chemical Enrichment History of Abell 1837}

\author[M. K. Erdim et al.]{
M. K. Erdim,$^{1,2}$\thanks{E-mail: mkiyami@yildiz.edu.tr (MKE)}
C. Ezer,$^{1}$
O. Ünver$^{1,3}$
F. Hazar$^{1}$
and M. Hudaverdi$^{1}$\thanks{E-mail:hudaverd@yildiz.edu.tr (MH)}
\\
$^{1}$Y{\i}ld{\i}z Technical University, Faculty of Science and Art, Department of Physics, Istanbul 34220, Turkey\\
$^{2}$Y{\i}ld{\i}z Technical University, Graduate School of Natural and Applied Sciences, Istanbul 34220, Turkey\\
$^{3}$Beykent University, Faculty of Engineering and Architecture, Istanbul 34396, Turkey\\
}

\date{Accepted XXX. Received YYY; in original form ZZZ}

\pubyear{2020}

\begin{document}
\label{firstpage}
\pagerange{\pageref{firstpage}--\pageref{lastpage}}
\maketitle

\begin{abstract}
In this paper, we report the relative SNe contributions on the metal budget of the ICM of Abell 1837 galaxy cluster at redshift z $=$ 0.069. For this purpose, we analysed the hot ICM of the cluster and obtained radial metal distributions by using XMM-Newton archival data with total exposure $\sim$100 ks. These metal measurements consist of Mg, Si, S, Fe and Ni within 0.7 R$_{500}$ radius which is divided into three concentric annuli. In order to explain the observed metal abundance pattern in terms of relative supernova contributions, we used our newly developed code \texttt{SNeRatio} which utilizes theoretical nucleosynthesis models. This study covers the most recent 3D SNIa and SNcc yield tables. All combinations of these theoretical yields were fitted with our measured abundance ratios and statistically acceptable ones were selected. Each of these models were found to predict a uniform SNIa percentage contribution to the total SNe from the cluster center to the outskirts and form an SNIa ratio distribution with a mean 39 $\pm$ 14$\%$. This uniformity is consistent with the early enrichment scenario which assumes that the metal production processes begin in early phase of cluster formation, namely proto-cluster phase at epoch z $\geq$ 2.

\end{abstract}




\section{Introduction}
Clusters of galaxies are remarkable probes of dynamical history and chemical enrichment since their evolution time is comparable to the age of the Universe. These massive objects can be portrayed by stars ($\sim$3$\%$) and X-ray emitting hot dilute gas ($\sim$12$\%$) trapped in deep dark matter ($\sim$85$\%$) potential well (for a review, see \cite{pratt2019galaxy}). Therefore, most of the baryonic matter resides in the form of hot gas constituting the so-called intra-cluster medium (ICM).

Observations reveal that the ICM is enriched in terms of metal abundance. Supernovae (SNe) explosions are regarded as main mechanisms to be responsible for creating this metal-rich environment. The main proposed transport mechanisms responsible for spreading these produced elements into the inter-galactic space are ram-pressure stripping \citep{Gunn1972}, galactic winds \citep{DeYoung1978} and Active Galactic Nuclei (AGN) outflows \citep[e.g.][]{Blanton2001,McNamara2001,Fabian2006,McNamara2007}. Core-collapse supernovae (SNcc) explosions account for both creation and dispersion of the $\alpha$ elements such as O, Ne and Mg; whereas Type Ia supernovae (SNIa) are responsible for the production and ejection of heavier metals such as Fe and Ni. Intermediate elements namely Si and S are presumably produced by both types. High resolution X-ray spectroscopy allows us 
to observe these metals via their emission lines which have been synthesised by
billions of SNe inside the cluster galaxies. A comparison of data and theoretical models, i.e SNe yield tables, 
is a well known method to obtain the fraction of each SNe types contributing to the enrichment.
This fraction provides important information about the chemical evolution of the Universe.

After the launch of ASCA, the metal abundances in the ICM were studied extensively \citep[e.g.][]{Mushotzky}. The spatially resolved metal measurements (e.g. Si, S, Fe) from center to larger radii for relaxed clusters of galaxies pointed toward an increasing SNcc dominance at clusters’ outskirts \citep{Finoguenov2000,Finoguenov2001}. These first results supported a scenario in which SNcc products are mostly produced before clustering and well mixed throughout the ICM, but SNIa products increase as a result of a late contribution in the central region after cluster collapse. The studies with better spectral resolution also verified similar enrichment processes, e.g. XMM-Newton observations of central regions {\citep{Bohringer2004,Tamura2004} and Chandra observations of region covering R$_{500}$\footnote{R$_{\Delta}$ is defined to be the radius at which mean density of the cluster is $\Delta$ times the critical density of the Universe at the cluster's redshift.}} \citep{Rasmussen2009}.

In time, however, late contribution of SNIa scenario has not been supported by many XMM-Newton studies from either European Photon Imaging Cameras (EPIC) or Reflection Grating Spectrometer (RGS) instruments. Observations covering only core regions reported decreasing or flat distribution of SNcc over SNIa products \citep{Werner2006, deplaa2006, Simionescu2009, Bulbul_2012, Mernier2015}. This phenomenon surely disagrees with SNIa dominance at the cluster's core, therefore requires investigation of the outskirts together with central regions to emerge ultimate enrichment history.  

Robust measurements on the clusters' outskirts were groundbreaking, and subsequently new insights have been gained about the history of chemical evolution of the ICM. Deep Suzaku observations ($\sim$ 1 Ms) of Perseus cluster unveiled uniform radial and spatial Fe distribution from $\sim$0.2 R$_{200}$ to R$_{200}$ \citep{Werner2013}. 
The value of $\sim$0.3 Solar Fe abundance in the outskirts supports that the metals were expelled from galaxies via galactic winds and AGN activities
at the major epoch of star formation (z $\simeq$ 2-3 \citep{Madau})
The results have also been confirmed by Fe-K line measurements from outskirts (r $>$ 0.25 R$_{200}$) of 10 nearby cool-core clusters \citep{Urban2017}. In the meantime, some studies focused on other elements along with Fe and quantified percentage contributions of types of SNe up to the virial radii. Suzaku Key Project observations of the Virgo Cluster revealed flat SNIa ratio profile out to 1.3 R$_{200}$ and reported that the value obtained as being consistent with the cluster cores and the Milky Way \citep{Simionescu2015}. Later, \citet{Mernier2017} examined this issue on a stacked analysis of 44 objects including galaxy clusters, groups, and elliptical galaxies from Chemical Evolution RGS Sample (CHEERS) project \citep{de2017cheers} and confirmed a similar uniform ratio profile up to 0.5 R$_{500}$. Finally, \citet{Ezer2017} also observed a uniform pattern through deep Suzaku observations of A3112 galaxy cluster from center to the virial radius.

The presence of a uniform radial distribution of SNIa/SNcc contributions up to the outskirts suggests that ICM enrichment occurred at early times and the metals were synthesized during or shortly before the formation of clusters and is thought to be the origin of the significant portion of the metals in the ICM that we observe today. The predictions from the simulations also show excellent agreement with the observations \citep{Planelles2014}. A uniform contribution of SNIa (SNcc) products in the ICM from simulated clusters stressed the importance of the AGN feedback for transport mechanism at star forming regions even beyond the virial radius \citep{Biffi2017,Biffi18}.

Some studies mentioned the uncertainties in the metal yields due to inadequacy of the theoretical models to explain the abundance pattern found in the ICM \citep{deplaa2007,Mernier2016}.
As a recent example, the abundance ratios in the central region of the Perseus cluster with the Hitomi (Astro-H) SXS instrument implied the necessity of neutrino physics in SNcc nucleosynthesis models \citep{Simionescu2019}.  
However, these searches were only limited to the central region of a single object, and detailed high energy resolution observations of more systems are still lacking to consolidate a global view on these issues. In this context, it is valuable to investigate the radial profiles of the SNe contributing to the enrichment and testing the yield models with different physical assumptions.

The currently operating X-ray imaging observatories has now spent more than two decades in orbit, and now we have sufficient number of clusters and groups with deep observations of archival data. Even moderate quality data at low redshifts provide wide spatial coverage up to the outermost regions and allows us to get further valuable insights on the current enrichment scenario. Therefore, in this study, we present the analysis result of dynamically relaxed galaxy cluster A1837 using XMM-Newton archival data. It is reported as a moderate cooling flow cluster \citep{White97} with a total mass of M$_{500}=1.482\times10^{14}$M$_\odot$ \citep{lovisari2019non}. At a redshift of z $ = $ 0.069 \citep{oegerle2001dynamics}, the whole EPIC field of view (FoV) allowing us to resolve metal abundances and derive accurate estimations from to core to the cluster outskirts up to 0.7 R$_{500}$. Using the radial abundance profiles we derived a radial fraction profile of SNe types and discussed the possible enrichment mechanisms responsible for releasing and mixing the metals observed in A1837.

This paper is structured as follows. The data reduction, background treatment and spectral fitting are described in Section \ref{sec:obs_red}. The yield calculation is given in Section \ref{sec:yieldcalculation}. 
We then present our newly developed code {\texttt{SNeRatio}} to generate SNIa fraction over the total number of SNe (SNIa+SNcc) in Section \ref{sec:sn_ratio}.
We introduce our temperature structure (Section \ref{results1}) and radial metallicity (Section \ref{results2}), as well as radial SNIa ratio profile (Section \ref{results:Sneratio}). We discuss and interpret the findings in Section \ref{sec:discussion}. 
We note that throughout this paper, we used  H$_{0}$ $=$ 70 km s$^{-1}$ Mpc$^{-1}$, $\Omega_{M}$ $=$ 0.30 and $\Omega_{\Lambda}$ $=$ 0.70. At the cluster's redshift, 1 arcsecond corresponds to 1.353 kpc. Unless otherwise stated, all errors are at $1\sigma$ confidence level.


\section{Observation and Data Reduction} 
\label{sec:obs_red}
For our analysis, we used the archival XMM-Newton \citep{jansen2001} observation of the central region of galaxy cluster A1837 (RA:$14^{h}01^{m}36.7^{s}$, DEC:$-11^{d}07^{m}28.0^{s}$) with 100.2 ks total exposure. The observation was performed January 11, 2001 with observation id 0109910101 and the thin optical blocking filter was used in Full Frame mode for the EPIC-MOS and Extended Full Frame mode for EPIC-pn.


\subsection{EPIC Analysis}
The raw data were processed with XMM-Newton Extended Source Analysis Software (ESAS), which is integrated into the Science Analysis System (SAS) version 18. In this paper, we followed ESAS analysis procedures based on the methods described by \citet{snowden2014cookbook}.

\begin{figure}
	\includegraphics[width=\columnwidth]{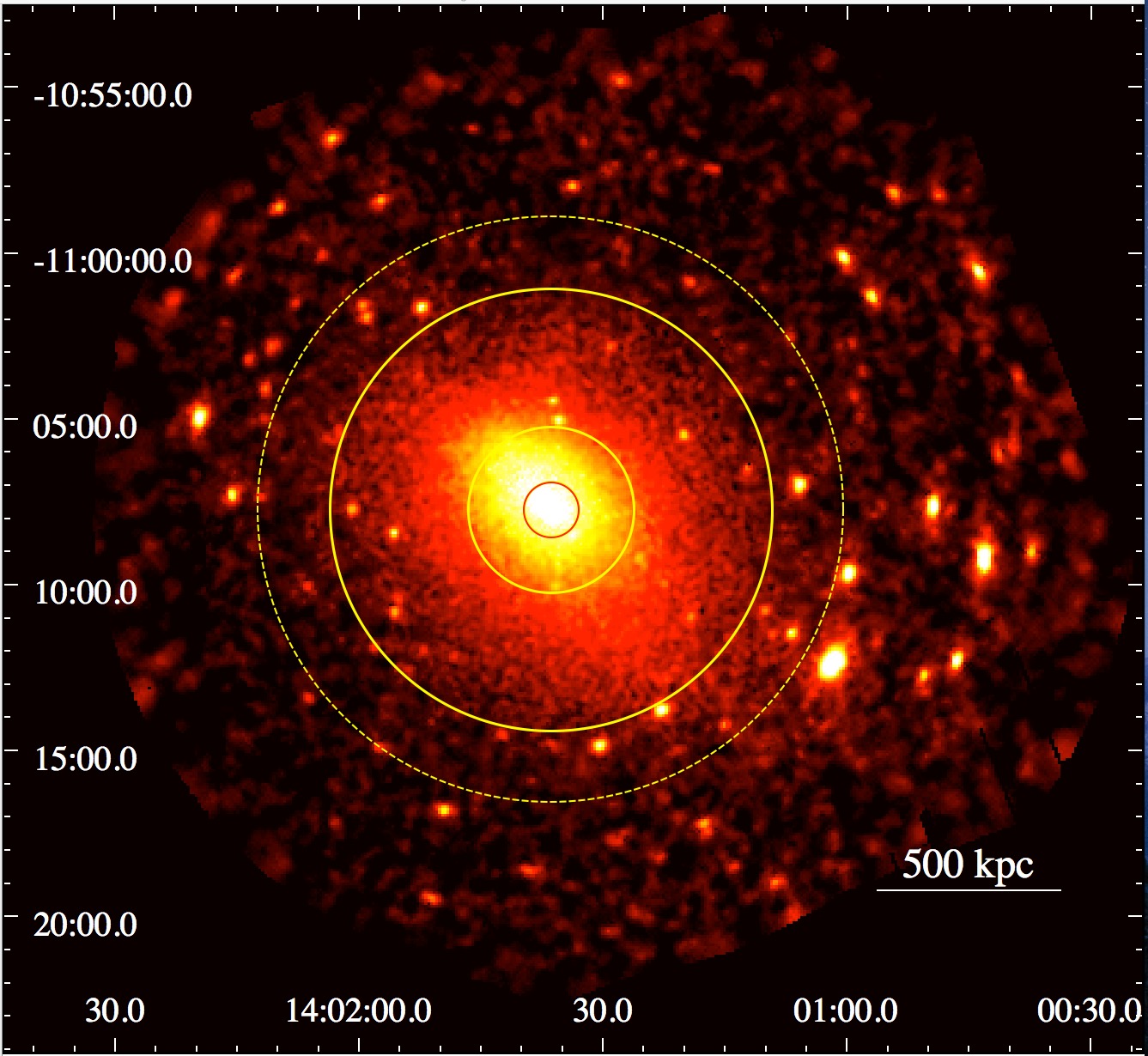}
    \caption{Adaptively smoothed EPIC combined image of A1837 in energy range of 0.4-7.2 keV. Dashed yellow circle shows the R$_{500}$ radius. The inner red circle and the yellow solid annular regions delimitate 0 $-$ 50$ '' $, 50$ '' $ $-$ 150$ '' $ and 150$ '' $ $-$ 400$ '' $ edges, respectively.}
    \label{Regions}
\end{figure}

The basic filtering and calibration were applied to MOS and pn event files using XMM-ESAS tools \textit{emchain} and \textit{epchain}. The event lists were filtered with "FLAG $== 0$" condition to reject bad pixels and columns. Single, double, triple and quadruple pixel events (PATTERN $\leq$ 12) for MOS and only double pixel events (PATTERN $\leq$ 4) for pn instruments were selected.
To remove the soft-proton (SP) contamination and generate the good time intervals (GTIs), \textit{mos-filter} and \textit{pn-filter} tasks were used. The GTI files were, then, used to determine the net exposures of the observation. In Table \ref{tab:obs_sum}, the observation summary and the net exposures are given for EPIC.

\begin{table}
\centering
\caption{Observation Summary of A1837}
\label{tab:obs_sum}
\begin{tabular}{@{}lcccc@{}}
\hline\hline
\multirow{2}{*}{Obs.ID}      & Date & Duration & Instrument & Net Exp.\\
		    &  (Start/End)    &  (ks)    &    (EPIC)    &      (ks)     \\\hline 
0109910101  & 2001-01-11/12 & 100.219  & MOS1 & $\sim $ 47.50 \\
		    &        &                 & MOS2 & $\sim $ 47.60 \\
		    &        &                 & pn  & $\sim $ 44.20    \\
\hline
\end{tabular}
\end{table}

After filtering the data, we used the \textit{Fin\_over\_Fout} public script \citep{de20042} provided by the XMM–Newton EPIC Background working group to check if additional cleaning is needed due to the residual SP contamination. We found the \textit{F$_{in}$/F$_{out}$} ratios as 1.054 $\pm$ 0.030 for MOS1, 1.037 $\pm$ 0.028 for MOS2 and 1.131 $\pm$ 0.032 for pn. The ratio smaller than 1.15 indicates that the filtered observations were clean enough and not contaminated by SP flares. Bright point sources were detected with ESAS task \textit{cheese}. 
We also visually corrected and reorganised the generated source list in order to minimise the false or missing detections.

For the spectral analysis, we divided the FoV into three concentric annular regions, centered at X-ray emission peak; 0 $-$ 50$''$ (0 $-$ 0.08 R$_{500}$), 50$''$ $-$ 150$''$ (0.08 $-$ 0.25 R$_{500}$) and 150$''$ $-$ 400$''$ (0.25 $-$ 0.7 R$_{500}$) edges respectively (see Fig. \ref{Regions}). The location of the X-ray peak was determined in detector coordinates by using the task \textit{xmmselect}. R$_{500}$ value of the cluster was adopted as R$_{500}$ $=$ 793 kpc from the study of \citet{piffaretti2005temperature}. Finally \textit{mos-spectra}, \textit{mos\_back}, \textit{pn-spectra} and \textit{pn\_back} tasks were used to obtain final spectral products which contain ancillary response file (ARF) and response matrix files (RMF) of selected regions for every EPIC instrument individually. 


\subsection{EPIC Background} \label{sec:epic_background}
The correct modeling of background radiation plays a crucial role to determine the basic properties such as temperature and metallicity of the ICM. Any missing component or inappropriate modeling of the background radiation may lead to incorrect scientific results \citep{deplaa2007}. For the EPIC background treatment, we followed the methods described in \citet{Snowden08,kuntz2008epic,snowden2014cookbook} and references therein. Background components mentioned in this study are the residual SP contamination, the quiescent particle background (QPB), the instrumental fluorescent lines and the cosmic X-ray background (CXB).

The residual SP contamination which affects all EPIC detectors was modeled by using a separate broken power law (bknpow) component with the diagonal response matrices supplied by XMM-ESAS CalDB. The break point, E$_{(break)}$, of the \textit{broken power law} was fixed to 3.0 keV. The lower limit of the photon index ($\Gamma_1$) was left free to vary within 0.1 $-$ 1.4 and the upper limit of the photon index ($\Gamma_2$) was allowed to reach up to 2.5 which is agreed with \citet{snowden2014cookbook}. While for each region the photon indexes of MOS were linked to each other, the photon index of pn was left free. The photon indexes for different regions were estimated in the physically tolerable range without having any inconsistency.  

The QPB spectra were extracted from the filter wheel closed (FWC) data by \textit{mos\_back} and \textit{pn\_back} tasks. The generated spectra were subtracted from the source spectra prior to fitting.
The instrumental fluorescent lines which dominate the spectrum at different energies were taken into account by using Gaussian profiles with zero-widths. The instrumental lines for MOS1 and MOS2 were Al K$\alpha$ ($\sim1.49$), Si K$\alpha$ ($\sim1.75$ keV). For the pn, we considered low energy Si K$\alpha$ ($\sim1.75$ keV), together with contribution from high energy lines Ni K$\alpha$ ($\sim$7.47 keV), Cu K$\alpha$ ($\sim$8.04 keV), Zn K$\alpha$ ($\sim$8.63 keV) and Cu K$\beta$ ($\sim$8.90 keV). We initially determined best-fit values of line energies and normalizations, then kept these lines fixed in our fits.

The CXB is spatially variable and composed of three components: one non-thermal component for the cosmic unresolved point sources (UPS), two thermal components due to Galactic Halo emission (GH) and the Local Hot Bubble (LHB) \citep{apec_2001}. We use the ROSAT All-Sky Survey (RASS) X-ray background tool to model the CXB provided by the High Energy Science Archive Research Web site\footnotemark{}. 
The annular region surrounding the cluster center between 1$^\circ$ and 2$^\circ$ was chosen to obtain the RASS spectrum. 
Abundance and redshift parameters of thermal components in the CXB model were fixed at 1.0 Z$_\odot$ and 0.0, respectively. 
The UPS contamination was modeled using an absorbed power-law with $\Gamma$ $=$ 1.45 \citep{Snowden08}. The best-fit parameters which were obtained from X-ray background tool summarized in Table \ref{tab:bkg_table}. Then, they were used as initial values while performing simultaneous fitting of source and RASS spectra. It was seen that the CXB parameter values are highly consistent with the studies of \citet{kuntz2000,Snowden08,wulf2019}.

\footnotetext[2]{\href{https://heasarc.gsfc.nasa.gov/cgi-bin/Tools/xraybg/xraybg.pl}{heasarc.gsfc.nasa.gov/cgi-bin/Tools/xraybg/xraybg.pl}}

{\renewcommand{\arraystretch}{1.2}
\begin{table}
	\centering
	\caption{The Cosmic X-ray background (RASS) fit parameters.}
\begin{threeparttable}

	\begin{tabular} {@{}lcc@{}}
		\hline\hline
		Component & Parameter & Value\\
		\hline 
		LHB & kT (keV) & 0.108\\
		    & Norm (cm$^{-5}$) & 3.957 $\times$ 10$^{-7}$\\ \hline
	    GH$_{Cold}$ & kT (keV) & 0.124\\
	        & Norm (cm$^{-5}$) & 3.398 $\times$ 10$^{-6}$\\ 
	    GH$_{Hot}$ & kT (keV) & 0.273\\
	        & Norm (cm$^{-5}$) & 1.551 $\times$ 10$^{-6}$\\ \hline
        UPS & $\Gamma$ & 1.45 \\ 
		    & Norm (cm$^{-5}$) & 9.034 $\times$ 10$^{-7}$\\ \hline
		    & N$_H$ (cm$^{-2}$)\tnote{*} & 4.31 $\times$ 10$^{20}$ \\
		\hline
\end{tabular}
\begin{tablenotes}
     \item[*] Fixed to the average value taken from \citet{ben2016hi4pi}.
\end{tablenotes}
\end{threeparttable}
\label{tab:bkg_table}
\end{table}}

\subsection{EPIC Spectral Fitting} 
\label{sec:fitting}
For the spectral fitting, we used X-ray Spectral Fitting Package XSPEC 12.11.1 \citep{xspec1999} and AtomDB version 3.0.9 \citep{apec_2001,foster2012updated}. MOS spectra were restricted in 0.5 $-$ 10.0 keV energy range whereas pn spectra in 0.6 $-$ 10.0 keV. The fits were performed using C-statistics \citep{Cash1979}. The elemental abundances were acquired with respect to the Solar \textit{ASPL} \citep{aspl2009} abundance table.

We modeled individually three spectra which were extracted from the annular regions 0 $-$ 50$^{\prime\prime}$, 50$''$ $-$ 150$''$ and 150$''$ $-$ 400$''$ centered at the cluster's X-ray peak. Besides that our selected regions have at least 6000 spectral counts. Since the spectra from MOS1, MOS2 and pn were fit simultaneously, two constants which represent the calibration offset and solid angles between the extraction regions of the EPIC instruments were added to the spectral model. The solid angle values were obtained from the \textit{proton$\_$scale} task and kept fixed during fitting. 

To investigate the physical properties of the ICM for all three regions, a variety of both absorbed single and multi-temperature models were taken into account. The final model selections with reasons and related abundance results were examined carefully in Section \ref{results1} and \ref{results2}.

\section{Yield Calculation}
\label{sec:yieldcalculation}

It is known that ICM is rich in metals and these metals are mainly synthesized by SNcc and SNIa explosions. To calculate the fractional contribution of individual SNe, we need a method comparing observed metal abundances with predicted nucleosynthesis yields from progenitor theories.  
These theories are simulations of supernova explosions that predict produced mass amounts for each element during an explosion according to definite initial conditions. These theoretical yields are generally given in Solar mass units (M$_\odot$). In the case of SNcc yields, they are frequently given with respect to a definite stellar mass, explosion energy, and metallicity. Thus, these yields should be integrated for each element by using an initial mass function (IMF), such as Salpeter \citep{salpeter} or top-heavy \citep{topheavy}. The integrated yields of a specified element $i$ are computed by;

\begin{equation}
   Y_i = \frac{\int_{m_l}^{m_u} Y_i(m)m^{-(1+x)}dm}{\int_{m_l}^{m_u} m^{-(1+x)}dm} 
\label{integratedyields}
\end{equation}

\noindent where $m_l$ and $m_u$ are lower and upper zero-age main sequence masses of stars expected to explode as SNcc in the final stage of their life. These values are usually set to $m_l = 10$ M$_\odot$ and  $m_u = 50$ M$_\odot$. For a standard Salpeter IMF $x = 1.35$, and for a top-heavy IMF $x=0.95$ due to the assumption of large massive star population \citep{salpeter,topheavy}. Throughout this study, SNcc yields were integrated with Salpeter IMF over the mass range of 10-50 M$_\odot$. In the case of SNIa, the yields do not need to be integrated over an IMF.

Abundance values for an element (X) with respect to Fe can be calculated with equation (\ref{singlemodel}) for a single SN type,

\begin{equation}
   <X/Fe> = \frac{Y_X/M_X}{Y_{Fe}/M_{Fe}} \times \frac{1}{(X/Fe)_\odot}
\label{singlemodel}
\end{equation}

\noindent where $Y$ and $M$ represent the yield and mass number (e.g. $M_{Fe} = 56$), respectively. Since the ICM is enriched by both SNe types, it is necessary to combine their fractional contributions as expressed in equation (\ref{multiplemodel}),

\begin{equation}
\label{multiplemodel}
  <X/Fe> = \left(\frac{a(N_{X})_{Ia}+b(N_X)_{cc}}{a(N_{Fe})_{Ia}+b(N_{Fe})_{cc}}\right) \times \frac{1}{(X/Fe)_\odot}\\
\end{equation}

\noindent where $N_x = Y_x/M_X$ and $a,b$ are the fractions of SNIa and SNcc contributions. Here, a and b should satisfy the following condition:

\begin{equation}
\label{SNIacc}
a+b =1.0.
\end{equation}

\section{SNeRatio Code} 
\label{sec:sn_ratio}
For this work and future purposes, we have developed an open source and well tested Python code called {\texttt{SNeRatio}} that is capable of predicting relative supernova contributions. This code uses given ICM abundances and fits them with a combination of selected progenitor yield models to calculate the relative contribution that explains the observed data. We briefly summarize the methodology in Section \ref{sec:yieldcalculation}.

{\renewcommand{\arraystretch}{1.2}
\begin{table}
\caption{Adopted Yields in {\texttt{SNeRatio}}}
\centering 
\begin{threeparttable}[b]
\begin{tabular}{@{}lccc@{}} 
\hline\hline 
Reference & Reference & Initial & SN \\
 &Table\tnote{a} &Metallicity\tnote{b} & Type\\
\hline 
\citet{Seitenzahl2013} & Table 2 & -- & SNIa\\\hline
\citet{Fink2013} & Table B1 & -- & SNIa\\\hline
\multirow{5}*{\citet{Nomoto13}} & \multirow{5}*{Table\tnote{c}}& 0.0 & \multirow{5}*{SNcc}  \\
& & 0.001 & \\
& & 0.004 & \\
& & 0.008 & \\
& & 0.02 & \\
& & 0.05 & \\ \hline
\end{tabular}
\begin{tablenotes}
     \item[a] This column refers to cited table numbers.
     \item[b] This column is given in units of Z$_{\odot}$
     \item[c] \href{http://star.herts.ac.uk/~chiaki/works/YIELD_CK13.DAT}{http://star.herts.ac.uk/chiaki/works/YIELDCK13.DAT}
\end{tablenotes}
\end{threeparttable}
\label{table:yieldmods} 
\end{table} }

The \texttt{SNeRatio} code includes a variety of SNe progenitor models (see Table \ref{table:yieldmods}). The elemental abundances measured from the X-ray spectra is compared with the implemented SN yields. Thereby, the SNIa fraction ($a$ value in equation (\ref{multiplemodel})) is estimated over the total number of SNe (SNIa+SNcc) contributing to the ICM enrichment. This fraction can be expressed as;

\begin{equation}
\label{RIa}
R_{(Ia)} = \frac{SNIa}{SNIa + SNcc}.   
\end{equation}

\noindent The code employs least-squares fitting algorithm and iteratively refines the best set of parameters.

The method was successfully applied to estimate the SNIa ratio on the chemical enrichment of A1837’s ICM, by offering the best progenitor model combination among the implemented options.The source code of the \texttt{SNeRatio} is publicly available from the webpage given below\footnote{\href{https://github.com/kiyami/sneratio}{https://github.com/kiyami/sneratio}}, along with the user manual. In addition to that, a direct access to web-app version of SNeRatio code with a well designed user interface is also provided in the link. Any contributions, feature requests and bug reports are all welcomed.

\section{Results} 
\label{sec:results}

\subsection{Temperature Structure}
\label{results1}

The conversion of equivalent width (EW) value to the related elemental abundance strongly depends on the plasma temperature.
In the presence of multi-phase gas, single temperature modelling of the ICM may underestimate \citep{Buote94,Boute2000} or overestimate \citep{Simionescu2009} metal abundances. 
As being a cool-core cluster, A1837 may host multi-phase gas in the central region ($<$ 0.08 R$_{500}$).
Therefore, defining thermal phase of the plasma is an essential step toward reliable measurements of the metal properties.
We fitted the central emission of the hot gas by two different absorbed temperature models: 
a single \textit{vapec} model \citep{apec_2001} in which the plasma is in collisional equilibrium and 
a multi-temperature \textit{vgadem} model in which the temperature has a Gaussian distribution.

In the fitting process, we set Mg, Si, S, Fe and Ni free, which were measurable in all three regions. The He abundance was fixed at 1.0 Z$_\odot$, and all other metals were linked to Fe. Hydrogen column density was fixed to its average LAB value reported in Table \ref {tab:bkg_table}.

Based on the \textit{vgadem} model, the temperature of the core is $3.47_{-0.07}^{+0.06}$ keV and the temperature width is $0.01_{-0.01}^{+0.42}$ keV consistent with zero. We found this temperature almost identical with $3.48_{-0.07}^{+0.07}$ keV obtained by single \textit{vapec} model which gives slightly better statistics. Adding another thermal component neither improved the fit, nor made the abundances constrained more tightly. 
Thus, we continue our analysis by modelling each spectra with single \textit{vapec} model.

Beyond the core, the temperature elevates in the adjacent area ($3.61_{-0.06}^{+0.05}$ keV) and then declines in the outskirts ($3.09_{-0.07}^{+0.07}$ keV). No steep gradients were observed within the radial temperature distribution. The spectral fitting results for three concentric annular regions can be found in Table \ref{table:vgaapec}.

{\renewcommand{\arraystretch}{1.5}
\begin{table*}
\begin{center}
\caption{Best-fit model parameters of abundances and temperatures for all three annuli, only the innermost region was fitted with \textit{vapec} and \textit{vgadem} models individually. The abundances were calculated by using ASPL Solar abundance table.}
\begin{tabular}{@{}ccc c c c c@{}} 
\hline\hline 
Region$^{\dagger}$ & \multicolumn{2}{c}{$R$ < 0.08}  && 0.08 < $R$ < 0.25 && 0.25 < $R$ < 0.70 \\
\hline
Model & \textit{vapec} & \textit{vgadem} && \textit{vapec} && \textit{vapec} \\
\hline \hline
kT (keV)& 3.48 $^{+0.07}_{-0.07}$ & 3.47$^{+0.06}_{-0.07}$ && 3.61 $^{+0.05}_{-0.06}$ && 3.09$^{+0.07}_{-0.07}$  \\ \hline
Mg & 0.25 $^{+0.26}_{-0.24}$ & 0.26$^{+0.27}_{-0.22}$ && 0.18 $^{+0.16}_{-0.14}$ && 0.34$^{+0.15}_{-0.14}$\\ \hline
Si  & 1.21 $^{+0.20}_{-0.20}$  & 1.21$^{+0.19}_{-0.16}$ &&  0.56 $^{+0.11}_{-0.10}$ && 0.40$^{+0.11}_{-0.10}$ \\ \hline 
S   & 0.88 $^{+0.28}_{-0.28}$ & 0.88$^{+0.26}_{-0.24}$ &&  0.44 $^{+0.16}_{-0.15}$ && 0.32$^{+0.16}_{-0.15}$ \\ \hline 
Fe  & 0.71 $^{+0.05}_{-0.05}$ & 0.72$^{+0.06}_{-0.03}$ && 0.48 $^{+0.03}_{-0.03}$ && 0.32$^{+0.03}_{-0.03}$ \\  \hline
Ni  & 1.63 $^{+0.56}_{-0.55}$ & 1.62$^{+0.50}_{-0.48}$ && 0.41 $^{+0.32}_{-0.31}$ && 0.45$^{+0.32}_{-0.30}$\\ \hline 
C-stat (d.o.f) & 3198.65 (3834) & 3198.69 (3833) && 4820.09 (5670) && 4210.86 (5669)\\
\hline 
\end{tabular}
\label{table:vgaapec} 
\end{center}
\footnotesize{$^{\dagger}$ The radii of the regions are in the units of $R/R_{500}$.} 
\end{table*}}


\subsection{Radial Metallicity Profile}
\label{results2}

\begin{figure}
	\includegraphics[width=\columnwidth]{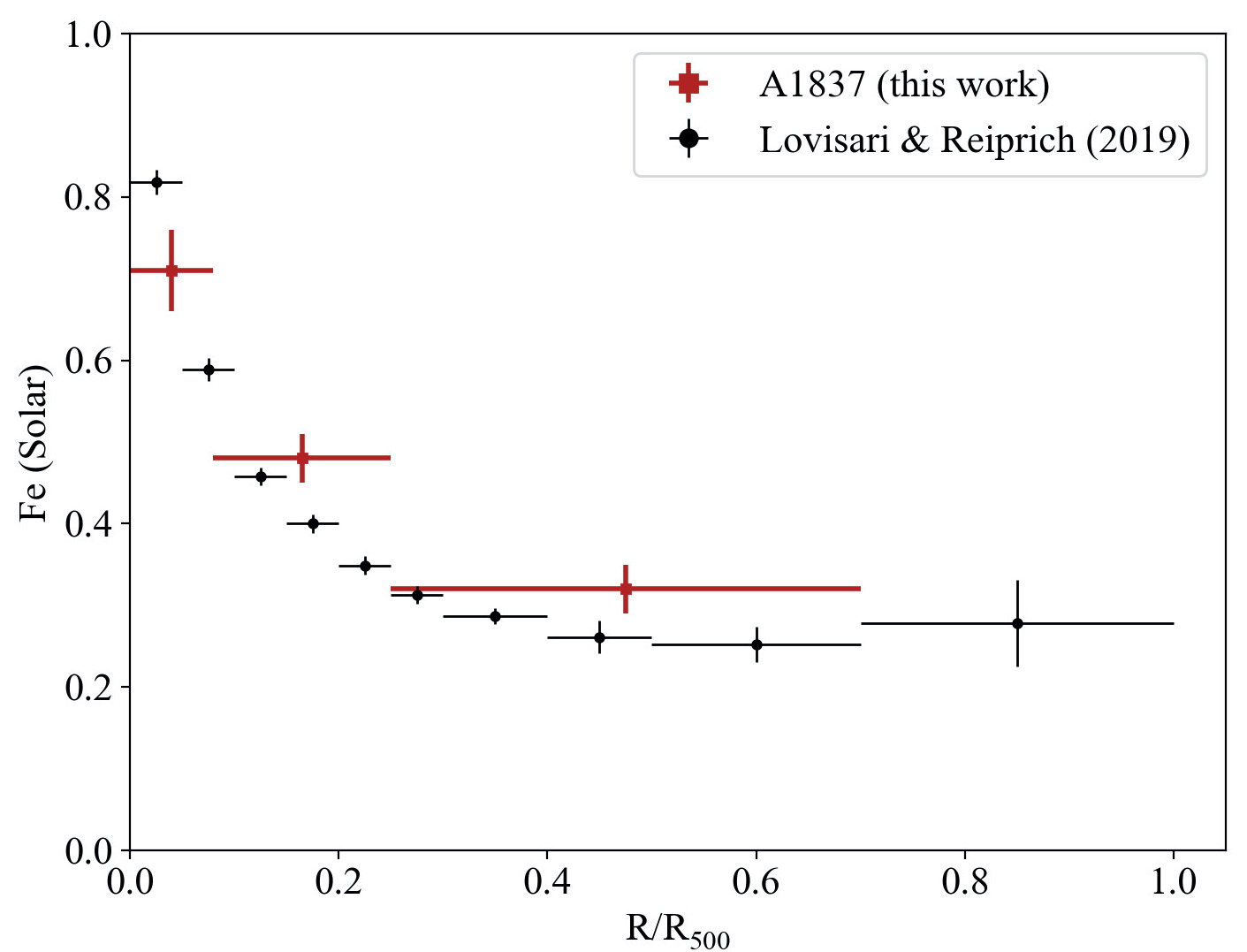}
    \caption{The red data points (boxes) represent the Fe profile of A1837 and the black data points (dots) show the average abundance profile of nearby relaxed systems reported by \citet{lovisari2019non}.}
    \label{fig:Fe_prof}
\end{figure}

\begin{figure}
	\includegraphics[width=\columnwidth]{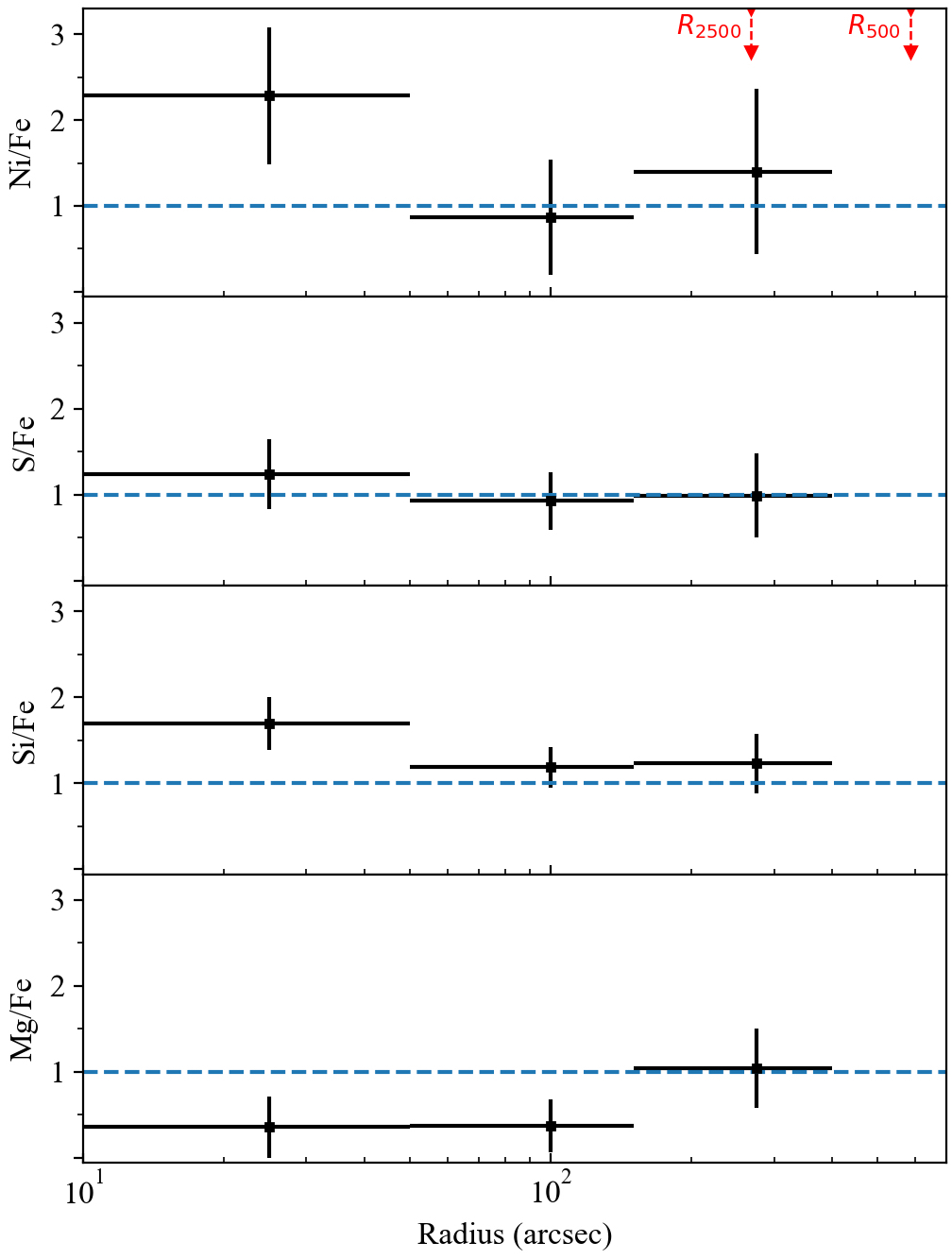}
    \caption{The radial profile of element abundances relative to Fe. All these elements were derived by using a single temperature \textit{vapec} model. The fractional errors were generated by adding in quadrature.}
    \label{Elements}
\end{figure}

We derived the radial distributions of specified metals throughout the ICM. Our Fe profile (Fig. \ref{fig:Fe_prof}) is found to be consistent with the average metallicity measurements of the relaxed galaxy clusters and groups reported by \citet{lovisari2019non}. In Fig. \ref{Elements}, we refer to the ratios of these elements with respect to Fe and these ratios were later used to quantify the percentage contribution of SNIa and SNcc processes to the total chemical enrichment of the ICM. Mg/Fe ratio decreases in the central regions, as being almost consistent with zero. Possible explanations and the effects of this situation in terms of the purpose of this work will be discussed in Section \ref{sec:discussion}.

\subsection{Radial SNIa Ratio Profile}
\label{results:Sneratio}

Spectral fits were performed on three concentric annular regions centered on the X-ray peak out to a radius of 0.7 R$_{500}$. The best-fit values of metal abundances are given in Table \ref{table:vgaapec}. Since the elements in the ICM are originated from SNIa and SNcc, we fitted this linear combination to our observed abundance pattern via \texttt{SNeRatio} code. We used the relative abundance values with respect to Fe to estimate the fraction of SNIa over the total SNe responsible for the enrichment (as in equation (\ref{RIa})).

During this fit we applied 3D SNIa models in near Chandrasekhar-mass, which are widely used in recent enrichment studies \citep{Mernier2017,Simionescu2019,mernier2020constraining}. These models consist of two different explosion mechanisms: delayed detonation from \citet{Seitenzahl2013} and pure deflagration from \citet{Fink2013}. We adopted SNcc yields with different initial metallicities from \citet{Nomoto13} (Z $=$ 0.0, 0.001, 0.004, 0.008, 0.02, 0.05). The references of these models are listed in Table \ref{table:yieldmods}.

\begin{figure*}
    \centering
	\includegraphics[width=0.98\textwidth]{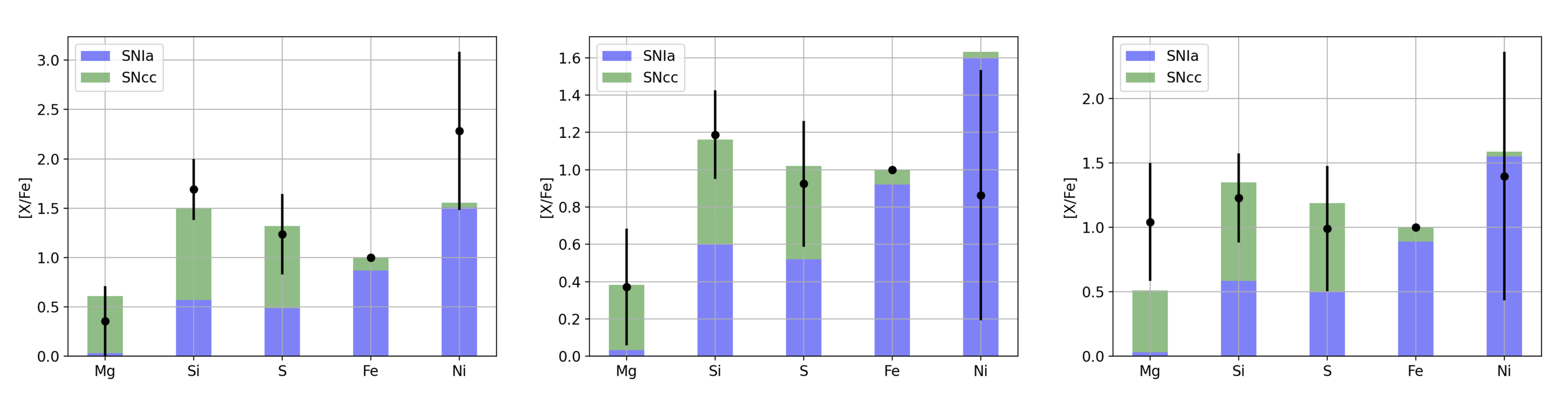}
    \caption{The relative contributions of SNIa (blue) and SNcc (green), adapting 3D spherically symmetric delayed detonation model N40 \protect\citep{Seitenzahl2013} and Z $=$ 0.0 with Salpeter IMF \protect\citep{Nomoto13} respectively. The outputs of the {\texttt{SNeRatio}} code for the innermost 0 $-$ 50$''$ ({\textit{left}}), 50$''$ $-$ 150$''$ ({\textit{middle}}) and  outmost 150$''$ $-$ 400$''$ ({\textit{right}}) annular regions.}
 \label{snratio_fit}
\end{figure*}

Each model indicates a different set of intrinsic physical assumptions. These different initial theoretical assumptions produce systematic difference for SNIa and SNcc type products, and thus the varying ratio. However, since the purpose of this work is to investigate SNe ratio profile from the core to the outskirt across the cluster, it would be more appropriate to represent the profile with the same physical model. In this way, the estimated values will be comparable with each region. Therefore, we need to determine a viable linear combination of SNIa and SNcc models which explains well all the regions simultaneously. In order to reduce the number of probabilities, the model combinations were left out of discussion in case the fit results $\chi^{2}_{\nu}$ > 2 in any region. Depending on this criterion, 112 model combinations out of 186 (6 models for SNcc and 31 models for SNIa) were selected. In this paper, we thus focus on those 112 acceptable combinations and discuss their possible implications.
 
\begin{figure}
	\includegraphics[width=1.0\columnwidth]{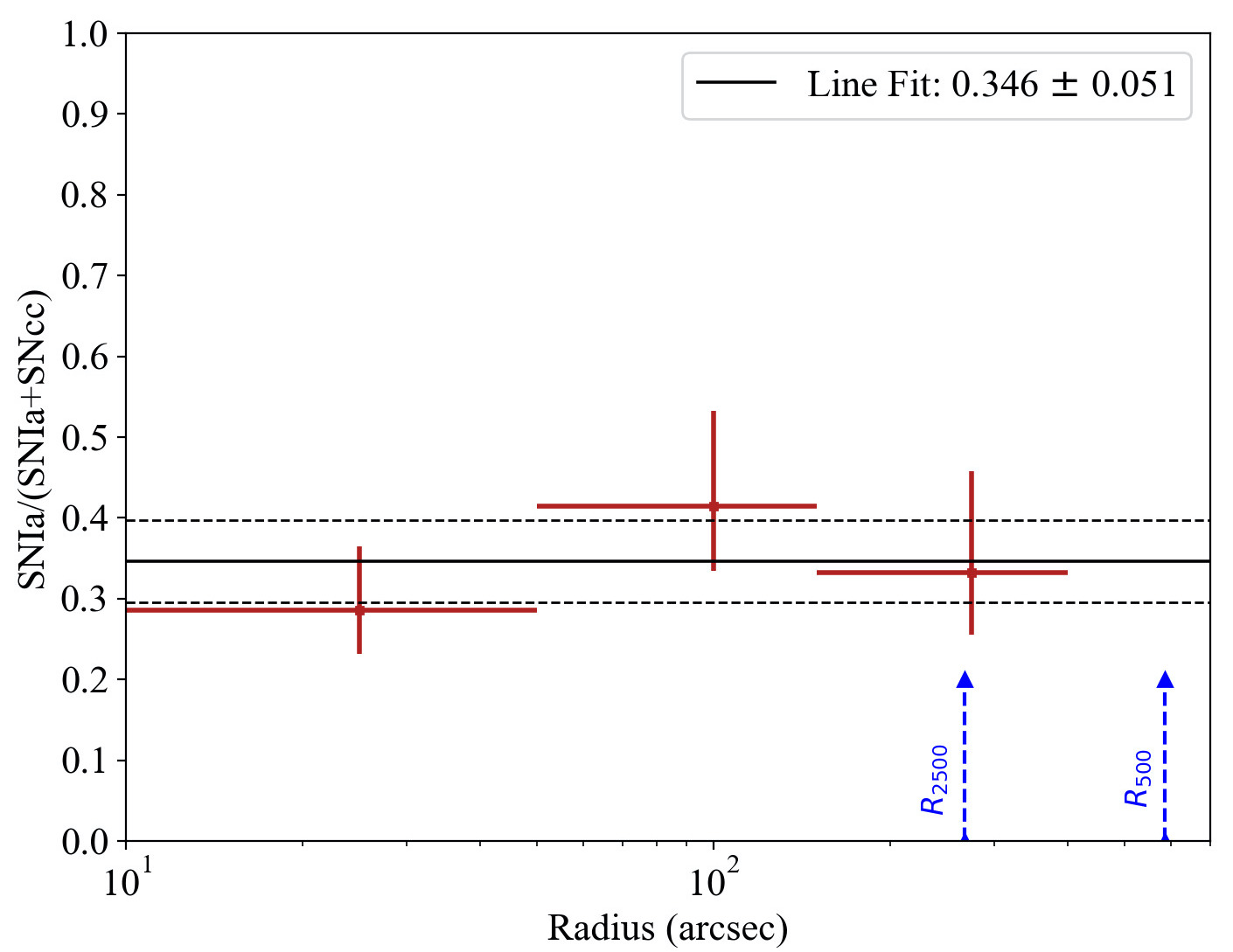}
    \caption{The radial profile of the SNIa ratio (red) obtained for the N40 and Z $ = 0.0 $ model combination. The black line represents 1$\sigma$ confidence interval of constant fit.}
    \label{N40}
\end{figure}

Before probing overall acceptable models, an optimal model combination for three annuli can be selected by minimising the sum of their $\chi^{2}_{\nu}$ values in quadrature. With this method, we found that the best reproduction of metals for all regions was achieved by using delayed detonation 3D N40 model for SNIa and initial metallicity Z $=$ 0.0 for SNcc. SNIa contributions to the total enrichment by this set of models were estimated as $28.6^{+7.9}_{-5.4}$ $\%$  with $\chi^{2}_{\nu}$ $=$ 0.59 for the inner region, $41.4^{+11.9}_{-8.0}$ $\%$ with $\chi^{2}_{\nu}$ $=$ 0.47 for second region and $33.2^{+12.6}_{-7.7}$ $\%$  with $\chi^{2}_{\nu}$ $=$ 0.56 for the outskirt. The graphical representations given by the {\texttt{SNeRatio}} code can be found in Fig. \ref{snratio_fit}. Although this model combination is selected as the best scenario generating the abundance pattern in all three regions, there are numerous combinations with almost the same fit statistics, thus the acceptable $\chi^{2}_{\nu}$ $<$ 2 constraint does not allow us to discriminate and favor a specific model. From there on, we also aim to approximate radial behaviour of SNIa fractions. As seen in Fig. \ref{N40}, SNIa fractions scatter smoothly around a mean value. The line fit gives almost zero slope (2.581$\times 10^{-4}$ with $\chi^{2}_{\nu}$ $=$ 1.174) and shows the fractional SNIa contribution to the total SNe has tendency to be constant from core to out-most region (0 - 0.7 R$_{500}$). Consequently, polynomial regression with a degree of zero, namely mean value of A1837 ratio distribution, gives 0.346 $\pm$ 0.051 with $\chi^{2}_{\nu}$ $=$ 0.656.

\begin{figure}
	\includegraphics[width=1.0\columnwidth]{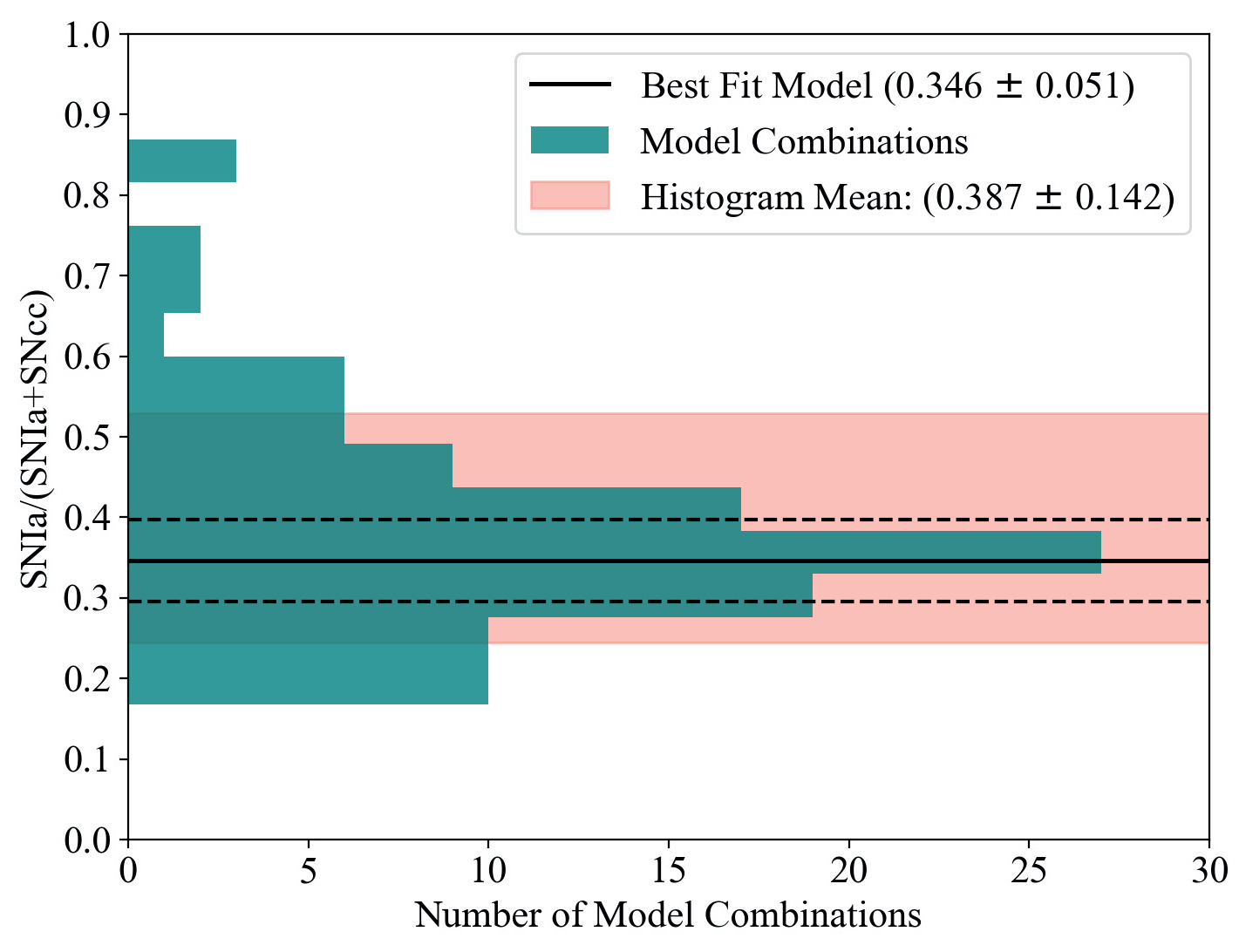}
    \caption{Histogram of SNIa ratios derived from line fits of 112 model combinations. The light-pink band shows the upper and lower limits of the mean value of the histogram, while the black line indicates the value of the best model combination (N40$+$Z$=$0.0).}
    \label{Numbermodelcomb}
\end{figure}

For all 112 acceptable model combinations, the radial distribution of SNIa ratio looks visually very similar to a flat distribution. After controlling the slope of each profile, we verified that the average radial slopes are in agreement with being flat with near zero slope (< 10$^{-3}$).  We determined the ratio values using fits to our profiles with a constant model. In particular, the uniform radial SNIa ratio to the total enrichment of ICM from core to outskirts is an important result. The ratio constant for 112 models varies widely from 0.1 to 0.9 around an average peak similar to a Gaussian distribution. The histogram in Fig. \ref{Numbermodelcomb} shows the number of model combinations versus the SNIa ratio. These ratios are somewhat dispersed, with a mean value of 0.387 $\pm$ 0.142. Our optimal model combination (N40$+$Z$=$0.0) coincides with the peak of the histogram. Thereby, it can be accepted as a representative model set for this distribution.

Finally, we made a comparison between SNIa models (delayed$-$detonation and deflagration) to investigate which one is more likely to reproduce our abundance pattern, and thus the SNIa ratio profile. Among 112 acceptable models, 68 of them are delayed$-$detonation and the remaining 44 are deflagration. These results do not highlight any of the explosion types significantly. We should also note that the mean of the ratio values predicted by the deflagration models (0.459 $\pm$ 0.158) was found to be slightly higher than the mean value predicted by the delayed$-$detonation models (0.340 $\pm$ 0.109).

\section{Discussion} 
\label{sec:discussion}
We investigated metal abundances of the galaxy cluster A1837 through the three concentric annular regions
centered on the X-ray peak out to a radius of 0.7 R$_{500}$. By using the observed abundance pattern and progenitor models in our newly developed \texttt{SNeRatio} code, we estimated the SNIa ratio over the total SNe necessary to enrich the ICM. The radial SNIa ratio distribution suggests a flat profile. Our main findings can be summarized as follows.

\begin{itemize}
    \item A1837 has a single-temperature phase plasma with a 3.48$\pm$0.07 keV value. The temperature reaches a maximum (3.61$_{-0.06}^{+0.05}$ keV) at 140 kpc. The central and outermost regions were observed to have mildly lower temperatures.
  
    \item The Fe profile clearly peaked towards the core, and agrees with the recent analysis by \citet{lovisari2019non} on average metallicities of the relaxed galaxy clusters and groups. In addition to that, our Fe abundances confirmed monotonically decreasing trend from the core to the outskirts, and at each bin the value drops by a factor of $\sim$ 1.5. 
    \item Since each element in the ICM is a mixture of synthesis from SNIa and SNcc, we selected a set of one SNIa and one SNcc model that reproduce the observed average abundance pattern. We have considered 6 SNcc models at different initial metallicities with Salpeter IMF integrated over 10M$_\odot$ $-$ 50M$_\odot$ and 31 near-Chandrasekhar mass 3D models for SNIa. The linear model combination was expected to fulfill $\chi_{\nu}^2$<2 condition while explaining the observed ICM metallicity pattern in each region. Results from this selection criterion indicate that:
       \begin{itemize}
        \item[--] 112 model combinations out of 186 satisfy the condition.  
        
        \item[--]Each model combination produced a flat radial distribution of SNIa ratio over the total cluster enrichment and found suitable to be represented by a constant. The uniformity of the ratio in the cluster is a strong evidence for early enrichment of the ICM, with most of the metals present being produced prior to clustering.
        
        \item[--] The SNIa ratio constants vary around a mean value of 0.39 $\pm$ 0.14 from 0.1 up to 0.9 (see histogram in Fig. \ref{Numbermodelcomb}) The fact that the ratio value scatters over such a wide range shows how the theoretical assumptions affect the calculations and how diverse they are. 
        
        \item[--] The SNIa ratio mean value is well consistent with statistically comprehensive state-of-art work reported as 29$-$45$\%$ for 44 cool-core clusters, groups, ellipticals by \citet{Mernier2016,Mernier2017}.

        \item[--] Similar studies observed uniform radial contributions of SNIa to the ICM enrichment. Their reported ratios from best-fit model combinations are 12$-$37$\%$ for Virgo \citep{Simionescu2015} and 12$-$16$\%$ for Abell 3112 \citep{Ezer2017}.  
        \end{itemize}

    \item The low Mg/Fe ratio in the central regions reported in this study is compatible with \citet{deplaa2006} and the same increasing trend from core to cluster's outskirts was also observed by \citet{Sakuma2011}. However, our Mg measurements below 0.25 R$_{500}$ might be questionable as the large error bars prevent us from deriving any firm conclusion.
    Relevant to this problem, significant differences on the Mg values were observed between three EPIC instruments individually, which were also previously reported in \citet{deplaa2007,Simionescu2009,Mernier2015}. In order to clarify whether this situation in Mg measurements causes bias in our SNIa ratio profiles:
     \begin{itemize}
     \item[--] We excluded Mg abundances and recalculated SNIa ratio to the total enrichment by using our \texttt{SNeRatio} code. We observed that this exclusion on each region did not change SNIa percentage contribution significantly.
     \item[--] Additionally, we also verified that this exclusion did not affect the flatness of the radial SNIa ratio profile. As an example, the observed change in the mean SNIa ratio was only 0.6$\%$ for SNIa N40 and SNcc Z=0.0 model combination. 
     \end{itemize}
    
    \item In respect to Ni/Fe inside the cluster's core ($<$ 0.08 R$_{500}$), our ratio (2.28 $\pm$ 1.42) is consistent within 1$\sigma$ confidence range with the value previously reported from investigation of 44 objects by XMM-Newton \citep{Mernier2016_I,Mernier2017}. However with the new version of atomic database SPEXACT(v3), re-calculated values of X/Fe are found to be around unity \citep{hitominife,Mernier18,Simionescu2019}, including Ni/Fe. Since our results are based on a different atomic database (APEC/ATOMDB version 3.0.9), we could not make a direct comparison. Nonetheless, \citet{aharonian2018atomic} compared their abundance values calculated by different versions of ATOMDB and SPEXACT. Their results point out a $\sim20\%$ drop in Ni/Fe ratio when switching from ATOMDB v3.0.8 to SPEXACT v3.00. Although we used ATOMDB v3.0.9 in our analysis, this study gives us a rough idea of how much change we would expect, if we used SPEXACT v3.00. 
   
\end{itemize}

It is also worth mentioning the results from short-lived Hitomi SXS \citep{hitominife} which revealed almost Solar abundance ratios for all measured elements with respect to Fe in the core of the Perseus cluster with great success.
However, none of the progenitor theories were found to be adequate enough to explain the observed abundance pattern \citep{Simionescu2019}. 
Although this result is based on a single object only, it implies that our current knowledge needs to be revised with future missions. Undoubtedly, the upcoming Athena X-ray Observatory \citep{nandra2013hot} is a promising candidate to improve the accuracy of the measurements. One of the science objectives undertaken by its payload X-Ray Integral Field Unit (X-IFU) is to reveal the chemical composition of the astrophysical objects at z $\leq$ 2 \citep{cucchetti2018athena,Barret2020,mernier2020constraining}. It also foresees to measure rare element lines which are important ingredients to understand the underlying mechanisms of SNe and their interrelated role in enrichment history.

\section*{Acknowledgements}
The authors thank the anonymous referee for constructive comments and suggestions. We would like to acknowledge financial support from the Scientific and Technological Research Council of Turkey (T\"{U}B\.{I}TAK) project number 118F035 and the EU COST Action CA16117 (ChETEC).

\section*{Data Availability}
The XMM-Newton raw data used in this article are available to download at the HEASARC Data Archive website\footnote{\href{https://heasarc.gsfc.nasa.gov/docs/archive.html}{https://heasarc.gsfc.nasa.gov/docs/archive.html}}. The reduced data underlying this article will be shared on reasonable request to the corresponding author.


\bibliographystyle{mnras}
\bibliography{example} 







\bsp	
\label{lastpage}
\end{document}